\documentclass[aps,prc,twocolumn,showkeys,showpacs]{revtex4-1}
\usepackage{amssymb}
\usepackage{graphicx,colordvi}

\begin{document}

\title{Beta decay of $^{252}$Cf on the way to scission from the exit point}

\author{K. Pomorski$^1$, B. Nerlo-Pomorska$^1$ and P. Quentin$^{2,3}$}
\affiliation{$^1$Katedra Fizyki Teoretycznej, Uniwersytet Marii
           Curie-Sk{\l}odowskiej, 20031 Lublin, Poland\\
       $^2$Universit\'e de Bordeaux, CENBG, UMR5797,33170 Gradignan, France\\
       $^3$CNRS, IN2P3, CENBG, UMR5797, 33170 Gradignan, France}

\date{\today}

\begin{abstract}
Upon increasing significantly the nuclear elongation, the  beta-decay energy
grows. This paper investigates within a simple yet partly microscopic
approach, the transition rate of the $\beta^-$ decay of the $^{252}$Cf nucleus
on the way to scission from the exit point for a spontaneous fission process. A
rather crude classical approximation is made for the corresponding damped
collective motion assumed to be one dimensional. Given these assumptions, we
only aim in this paper at providing the order of magnitudes of such a
phenomenon. At each deformation the energy available for $\beta^-$ decay, is
determined from such a dynamical treatment. Then, for a given elongation,
transition rates for the allowed (Fermi)  $0^+ \longrightarrow 0^+$ beta decay
are calculated from pair correlated wave functions obtained within a
macroscopic-microscopic approach and then integrated over the time corresponding
to the whole descent from exit to scission. The  results are presented as a
function of the damping factor (inverse of the characteristic damping time) in
use in our classical dynamical approach. For instance, in the case of a descent
time from the exit to the scission points of about $10^{- 20}$ second, one finds
a total rate of beta decay corresponding roughly to 20 events per year and per
milligram of  $^{252}$Cf. The inclusion of pairing correlations does not affect
much these results.
\end{abstract}
\maketitle

\section{Purpose}
\begin{figure}[b]
\includegraphics[width=1\columnwidth]{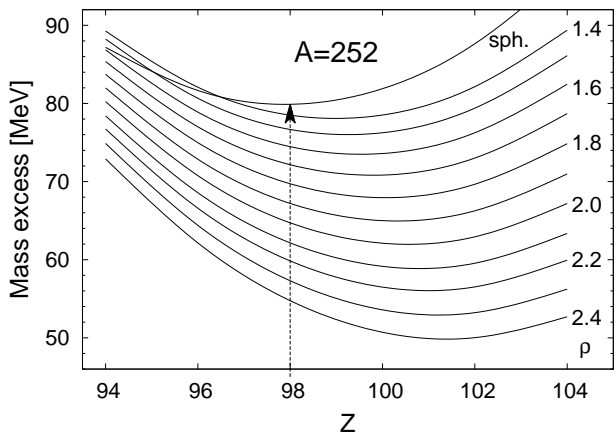}
\caption{Mass excess of Pu - Ku nuclei with A=252 evaluated within the LSD
model \cite{LSD}. Each curve corresponds to different elongation of nuclei
$\rho$ defined in the text. These values are reported on the r.h.s. side of
the every second curve. The upper curve (labelled as "sph.") presents the mass
excesses calculated for spherical nuclei.}
\label{fig1}
\end{figure}

Due to the decrease of the Coulomb energy and the increase of the surface
energy, upon strongly deforming a nucleus, it may happen that a nucleus which is
$\beta^-$-stable at equilibrium deformation gets a positive
$Q(\beta^-)=(M(Z,A)-M(Z+1,A)-m_e) c^2$ at large deformation. This is exemplified
on Fig.~\ref{fig1} within a Liquid Drop Model (LDM) approach \cite{LSD}, where
the masses of $A = 252$ isobars (from Plutonium to Kurchatovium) are plotted for
various deformations (defined as the ratio $\rho=R_{12}/R_0$ of the distance
$R_{12}$ between the mass centre of nascent fragments and the radius $R_0$ of
the corresponding spherical nucleus). One sees clearly that the beta stability
valley shifts from Californium almost to Nobelium when  the nuclear shape varies
from a sphere to configurations close to the scission point where one has
typically in such very heavy nuclei $\rho \approx 2.5$.

This result could be also attained upon considering the opposite behavior of
proton and neutron separation energies as could be seen from an extremely crude
liquid drop approach. Indeed, let us consider a Liquid Drop energy formula
comprising merely volume, surface, direct Coulomb and volume symmetry terms. For
a spherical shape, one gets for the (positive) separation energies for protons
($S_p$) and neutrons ($S_n$) with a usual notation and approximating the energy
differences as differentiable quantities (i.e. making a continuous approximation
for the variables $Z$ or $N$):
\begin{equation}
\begin{array}{l}
S_p = S^{(0)}_p + \frac{1}{3 } a_c \hspace{2mm}Z^2 A^{-\frac{4}{3}} -\frac{2  }{
3 } a_s \hspace{2mm}A^{-\frac{1}{3}} - 2 a_c \hspace{2mm}ZA^{- \frac{1}{3}}\\
S_n = S^{(0)}_n + \frac{1}{3 } a_c \hspace{2mm}Z^2 A^{-\frac{4}{3}} -\frac{2  }{
3 } a_s \hspace{2mm}A^{-\frac{1}{3}}
\end{array}
\end{equation}
where $S^{(0)}_p$ and $S^{(0)}_n$ are deformation independent terms given e.g.
for the protons by
\begin{equation}
S^{(0)}_p = - a_v + a_{sym.}\hspace{2mm} \frac{(N-Z) (Z+3N)}{A^2}
\end{equation}
\begin{figure}[htb]
\includegraphics[width=1.1\columnwidth]{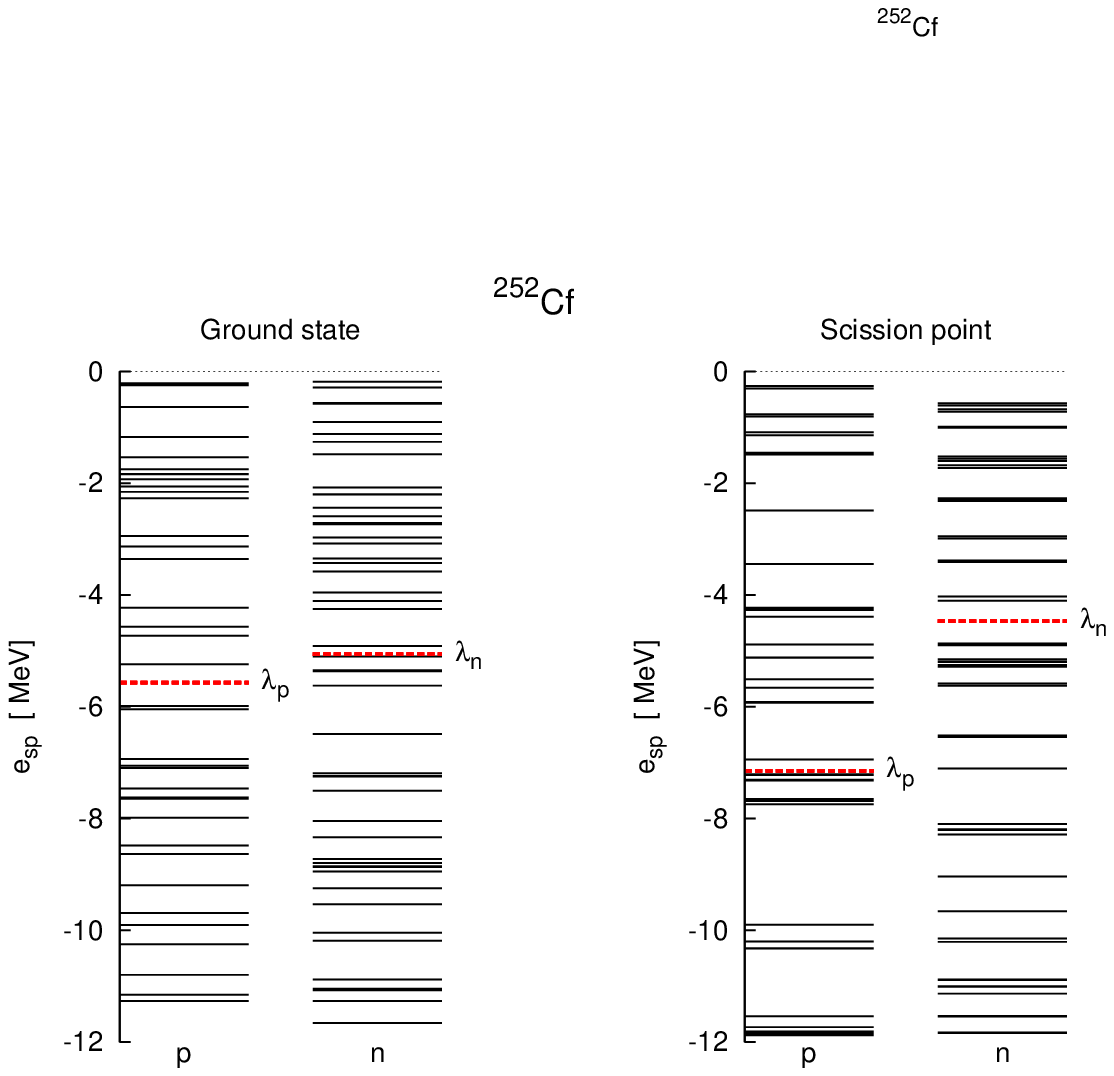}
\caption{Proton (p) and neutron (n) single-particle levels in the ground-state
(l.h.s.) and at the scission point (r.h.s.).}
\label{fig2}
\end{figure}

It appears clearly that upon deforming the nuclear shape, both the surface and
Coulomb deformation-dependent terms present in $S_n$ will contribute to its
decrease. These terms are also present in $S_p$ but are supplemented by a term
which increases with the deformation. We thus conclude that the difference $S_p
- S_n $  is increasing with deformation from about zero for the ground state of
a $\beta$ stable nucleus to a value which could be of the order of some MeV.
Resorting now to state of the art macroscopic-microscopic calculations (using
the LDM model of Ref. \cite{LSD} and the Yukawa folded potential of Ref.
\cite{YFP}) including pairing correlations using a seniority pairing force, one
gets an inversion of the relative position of neutron and proton Fermi levels of
the $^{252}$Cf beta stable nucleus upon increasing the elongation, the latter
becoming larger than the former. This is illustrated on Fig.~\ref{fig2} showing
the neutron and proton single particle (s.p.) spectra around the Fermi levels
for $\rho \approx 0.9 $ and $\rho \approx 2.8$ corresponding respectively to the
ground state and the scission point solutions.

We are therefore considering the following $\beta^{-}$ decay during the
spontaneous fission process (beyond the exit point):
\begin{equation}
^{252}{\rm Cf}_{154} \longrightarrow ^{252}{\rm Es}_{153} + e^- + \bar{\nu}
\end{equation}

We limit ourselves to allowed transitions. The spin and parity of the compound
nucleus being conserved during the fission process (in the absence of particle
emission) we are considering thus a  $0^+ \longrightarrow 0^+$ transition. This
rules out a Gamow-Teller transition and we are thus simply left with a Fermi
transition.

\section{Approximations}
\begin{figure}[htb]
\includegraphics[width=1\columnwidth]{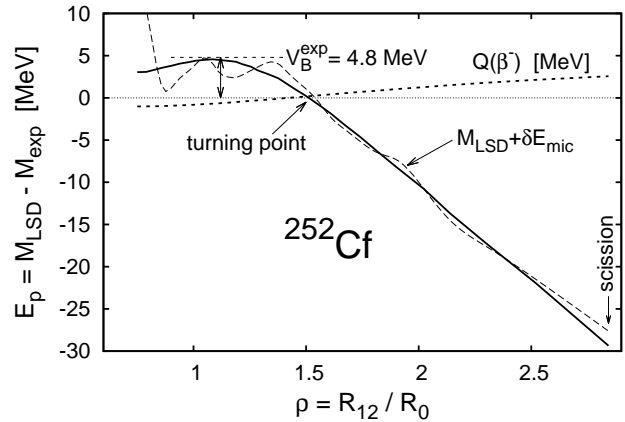}
\caption{ Macroscopic (solid line) and macroscopic-microscopic (dashed
line) fission barriers and the beta-decay energy $Q_\beta$ (dotted line) of
$^{252}$Cf as a function of nuclear elongation.}
\label{fig3}
\end{figure}

At this very preliminary stage of our study, we make the following
approximations pertaining to the static and dynamical calculations as well as to
the evaluation of transition rates.

a) We consider only a single fission path assumed to be the most probable one
out of static macroscopic - microscopic calculations of the above discussed
type. Pairing is included, as it should, to define adequately this optimal path.
On Fig.~\ref{fig3}, we have plotted the corresponding fission barrier down to
the scission point, along with the underlying LDM one.

As  a result we find that apart from the ground state region, and in particular
at large deformations (namely $R_{12}/{R_0} > 1.5)$, the LDM energy curve
represents a reasonable approximation of the results, where shell and pairing
effects are taken into account. However, for the calculated ground state one has
definitely to make a choice about its exact definition. One may rule out the LDM
solution since shell and pairing effects increase its height. As a good
approximation of the macroscopic-microscopic exit point, as can be checked on
Fig.~\ref{fig3}, is the exit point on the LDM fission
barrier (on the side of the descent towards fission) which is located at exactly
4.8 MeV below the saddle point. This value corresponds to the experimental
fission barrier height (noting en passant that our macroscopic-microscopic
calculations do reproduce it quite well).

In what follows we will use also, instead of the ratio $\rho = R_{12}/{R_0}$, a
collective variable $q$ obtained by a translation of $\rho$ such that $q = 0$ at
the exit point. We define in a somewhat arbitrary fashion a single scission
point. It corresponds to the point on the LDM fission path, where the nuclear
liquid drop ceases to be consisting of one single piece (it is obtained here for
$\rho \approx 2.9$). It corresponds to a gain in energy of $\Delta E_{max}
\approx 28$ MeV with respect to the exit point energy (or the ground state one,
since they are of course the same).

On Fig.~\ref{fig3} we have also plotted the LDM $Q(\beta^-)$ value as a
function
of $\rho$. It varies from  -1.2 MeV at sphericity to enter the instability
region for $\rho \approx 1.5$ which happens coincidentally, to be very close to
the exit point to reach $\approx 2.5$ MeV around the scission point.

b) In the present approach, which aims only at providing orders of magnitude,
we make the following approximations for the collective dynamics:\\
- As a function of the collective coordinate $q$, one may approximate
reasonably
(see Fig.~\ref{fig3}) the potential energy ${E_p}(q)$ as a linearly decreasing
function when $q$ varies from zero to $q_s$ which the value of this parameter at
scission. One has thus
\begin{equation}
{E_p}(q) = - \alpha q \hspace{10mm}\alpha\approx 20 \hspace{10mm} (0 \leq q
\leq q_s)
\end{equation}
- We approximate the mass parameter $m(q)$ as being constant $m(q) = \mu$ during
the descent from the exit point to scission and equal to the reduced mass
between the two final fragments (to assess the relevance of this approximation
see Ref. \cite{MASS}).\\
- We assume that the damping of the motion results from a simple friction force
$ - k \dot q$ with $k > 0$.

Then the equation of motion yields the following trivial solution (introducing
of course a vanishing collective velocity at the exit point)
\begin{equation}
q = \frac{\alpha}{k}\lbrack t - \frac{\mu}{k}(1 - e^{-
\frac{kt}{\mu}})\rbrack\,\,,
\label{qt}
\end{equation}
which could be numerically inverted to get the time $t(q)$ as a function of the
deformation $q$.

Then the collective kinetic energy will be given by
\begin{equation}
{E_K}(q) = \frac{1}{2}\hspace{2mm} \frac{\mu\alpha^2}{k^2}(1 - e^{-
\frac{kt}{\mu}})^2\,\,.
\end{equation}
The available energy is
\begin{equation}
\Delta E(q)= - {E_p}(q) = \alpha q \,\,,
\end{equation}
from which one gets the excitation energy ${E^\star}(q)$ as
\begin{equation}
{E^\star}(q) = \frac{\mu \alpha^2}{k^2}\lbrack x - g(x) -
                \frac{1}{2} g(x)^2\rbrack \,\,,
\end{equation}
where the dimensionless quantity $x$ is defined by
\begin{equation}
x = \frac{k t}{\mu}
\end{equation}
and with
\begin{equation}
g(x) = 1 - e^{-x} \,\,.
\end{equation}

c) To compute beta decay rates, the effect of pairing correlations will be taken
into account. For the sake of clarity we will first discuss our approach without
them and in a separate section the modifications brought in by their inclusion.

The conservation of the axial symmetry is assumed along the path. Each s.p.
state is thus defined in particular by the usual angular momentum projection $K$
quantum number.

We will assume that there is no polarization i.e. that the s.p. wavefunctions
in
the $^{252}$Cf or $^{252}$Es mean fields are the same and furthermore that one
has, with a transparent notation, the following relation between the binding
energies of the parent and daughter nuclei (Koopmans approximation)
\begin{equation}
E(^{252}{\rm Es}) = E(^{252}{\rm Cf}) + e_p - e_n \,\,.
\end{equation}

\section{Effective $Q_{\beta}$ values and available phase-space}

At a given elongation $q$ or at given time $t(q)$, the parent nucleus has an
excitation energy of $E^{\star}(q)$. The effective $Q_{\beta}(q)$ value for a
transition from a neutron single particle state $i$ to a proton single particle
state $f$ is thus given by
\begin{equation}
Q_{\beta}^{(i,f)}(q) = E^{\star}(q) + e_n(i) - e_p(f)) + 0.78 \ {\rm MeV}\,\,,
\end{equation}
where the last constant origins from the mass difference of neutron and proton
plus the electron mass.
Defining the neutron Fermi level $\lambda_n$ and similarly for protons
$\lambda_p$, one will consider an effective neutron Fermi level as
$\lambda^{\rm eff}_n(q)$
as
\begin{equation}
\lambda^{\rm eff}_n(q) = \lambda_n + E^{\star}(q) + 0.78 \ {\rm MeV} \,\,.
\end{equation}
Possible transitions between neutron s.p. states and proton s.p. states will
require the following conditions to be satisfied
\begin{equation}
\begin{array}{l}
K_n(i) = K_p(f) \,\,,\\ \lambda_p \leq e_p(f) \leq e_n(i) \leq
\lambda^{\rm eff}_n(q)\,\,.
\end{array}
\end{equation}

\section{Transitions rates}

The nuclear matrix element for such a $(i,f)$  Fermi transition is given with a
usual notation by
\begin{equation}
M_{i,f}=\int \!\!\!\int \!\!\!\int \phi_{p(f)}^{\star}(\vec{r})
\phi_{n(i)}(\vec{r}) d^3 r \,\,.
\end{equation}
A special note is to be made here, about the intrinsic parity breaking. Let us
call $| \Psi \rangle$ the intrinsic wavefunction breaking the left-right
symmetry, either for the parent or the daughter states. One should project a
positive parity state out of it, as
\begin{equation}
 | \Psi^{(+)} \rangle| = \frac{1}{\sqrt{2}} (|\Psi \rangle| + \hat{\Pi} |\Psi
\rangle) \,\,,
\end{equation}
where $ \hat{\Pi} $ is the parity operator.

The above nuclear matrix element will thus comprise two parts corresponding to
the two overlaps $\langle \Psi(i)| \Psi(f) \rangle$ and $\langle \Psi(i)|
\hat{\Pi} |\Psi(f) \rangle$. However it is known from parity projection
calculations of the fission barrier of heavy nuclei, see Ref. \cite{HAO}, that
somewhere before the second fission barrier and beyond, the intrinsic parity
breaking deformation is so large that the overlap between the corresponding
wavefunction and its parity image is negligible. We are just left here with the
$\langle \Psi(i)| \Psi(f) \rangle$ overlap as in the non-parity breaking case.
Yet, in this case one cannot, of course, apply a selection rule on the quantum
number $\pi $ which is not conserved but only on the angular momentum
projection $K$ which we have assumed to be a good quantum number.

The associated transition rate is \cite{WUM}
\begin{equation}
R_{i,f} = G^2 \hspace{2mm}\frac {| M_{i,f}|^2 \hspace{2mm}m_e^5 c^4}{2\pi^3
\hspace{2mm}\hbar^7} \hspace{2mm} f(Z=99,Q^{i,f}_{\beta}) \,\,,
\end{equation}
which after inserting of the week interaction coupling constant ($G$) and other
constants
values takes the following form:
\begin{equation}
R_{i,f} = 1.105 \hspace{2mm}10^{-4}\hspace{2mm}| M_{i,f}|^2\hspace{2mm}
f(Z=99,Q^{i,f}_{\beta}) \,\,.
\end{equation}
A rough estimate of the function $f$ as a function of the effective Q value
$Q_{\beta}^{(i,f)}(q)$ is given for $Z = 99$ by
\begin{equation}
\log_{10}(f) = 3.5\hspace{1mm} \log_{10}(Q_{\beta}^{(i,f)}(q)) + 3 \,\,,
\end{equation}
where $Q_{\beta}^{(i,f)}(q)$ is expressed in MeV.

\section{Including pair correlated nuclear states}

When pairing correlations are included in both the parent and daughter nuclear
states (assuming again no polarization effects, i.e. taking the s.p.
wavefunctions, occupation probabilities and quasi-particle energies as obtained
in the parent nucleus) one has just to multiply the overlaps $ M_{i,f} $ by the
BCS factor $u_{p(f)}v_{n(i)}$ while the effective $Q_{\beta}^{i,f}(q)$ becomes
\begin{equation}
\begin{array}{ll}
Q_{\beta}^{(i,f)}(q) &= E^{\star}(q) - [E^{qp}_n(i) + E^{qp}_p(f)] +
(\lambda_n - \lambda_p)\\ &+ 0.78 \ {\rm MeV} \,\,,
\end{array}
\end{equation}
where $E^{qp}_n(i)$ and $E^{qp}_p(f)$ are the initial neutron and the final
proton quasiparticle energies.

In this case, one may consider all transitions from single neutron to single
proton states which satisfy the following conditions beyond the angular
momentum component matching rule ($K_n(i) = K_p(f)$):\\
- the proton state should not be fully occupied (i.e. lying below the valence
space),\\
- the neutron state should not be fully unoccupied (i.e. lying above the
valence space),\\
- the final state energy should be lower or equal to the initial state energy,
i.e. $Q_{\beta}^{i,f}(q) \geq 0$ .

\begin{figure}[htb]
\includegraphics[width=1\columnwidth]{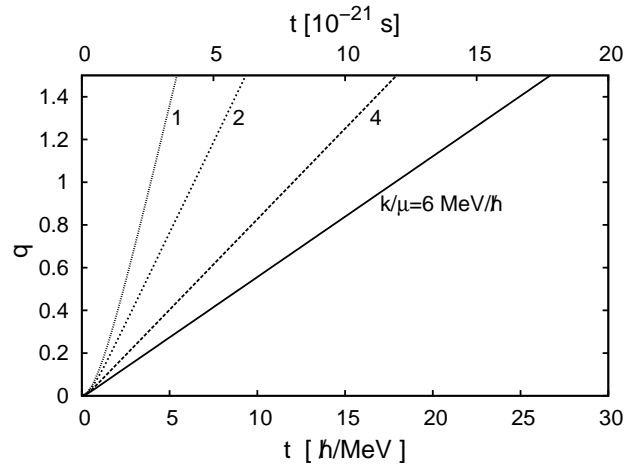}
\caption{Collective elongation $q=(R_{12}- R_{12}^{\rm exit})/R_0$ as
function of time (eq.~\ref{qt}) for different values of the nuclear damping
$k/\mu$.}
\label{fig4}
\end{figure}

\section{Total transition rates}

At a given time (or at a given elongation) the total rate of decay will be
obtained by summing all individual rates $R_{i,f}$ to get $R(q)$. For the whole
descent from exit to scission, one will get the number of decay for a single
fission, i.e. the probability of decay, by integrating $R(q)$ over the whole
time
of descent from the exit to the scission points
\begin{equation}
P_{decay} = \int_{0}^{q _{s}} R(q) \hspace{2mm}\frac{k}{\alpha}\hspace{1mm}
{(1-e^{- \frac{kt}{\mu}})}^{-1}dq \,\,.
\end{equation}
This probability, and similarly the pre-scission kinetic energy $E_K^0 =
{E_K}(q_s)$, are of course strongly dependent on the retained value for the
friction parameter $k$. In the following we will use our estimates of
$P_{decay}$ as a function of the time $t_{sc}$ to evaluate the number of 
$\beta$-decays from a sample of $^{252}$Cf in a given time period. 

\section{Results}
\begin{figure}[h!]
\includegraphics[width=1\columnwidth]{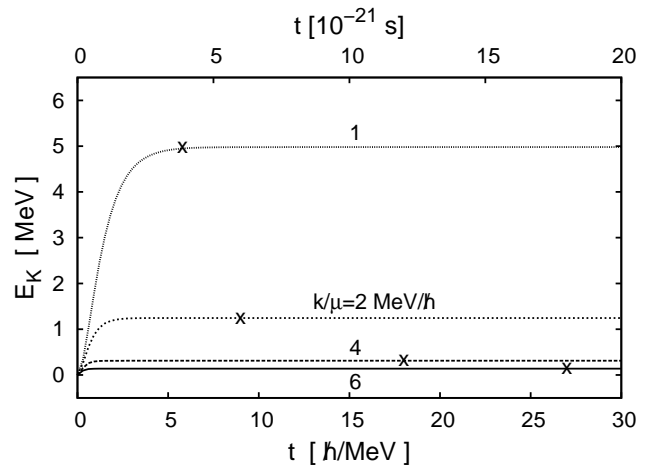}
\caption{Pre-scission kinetic energy for a few values of the reduced friction
as function of time. The time at which scission configuration is reached for a
given $k/\mu$ value is marked by a cross.}
\label{fig5}
\end{figure}

Let us first use the results of our simple dynamical calculations to
describe the collective motion from the exit to the scission points. Our results
will be presented for a few values of the reduced friction parameter $k/\mu$,
which inverse is the damping time (\ref{qt}). The damping parameters $k/\mu$ 
is given in units MeV$/\hbar \approx 1.5\,\,10^{21}{\rm s}^{-1}$.

As seen in Fig.~\ref{fig4}, the collective descent from $q = 0$ to $q \approx
1.4$ takes from {$\approx 3.5 \hspace{1mm}  10^{-21} s$ for a large dumping time
($k/\mu = 1$ in the above discussed units) to {$\approx 18 \hspace{1mm}
10^{-21} s$ for a small damping time ($k/\mu = 6$)). In all considered cases
the collective motion is completely damped at scission as exemplified on
Fig.~\ref{fig5}. The pre-scission kinetic energy $E_K$ values range from a
couple hundred keV for the large damping case ($k/\mu = 6$ ) to {$\approx 5$
MeV for the low damping case ($k/\mu = 1$).

The resulting number of beta decays for a 1 mg sample of $^{252}$Cf, is given in
Fig.~\ref{fig6}, as a function of the descent time $t_{sc}$ from the exit to the
scission points (or equivalently as a function of the damping parameter
$k/\mu$). It ranges from $\approx 6$ for the low damping case ($k/\mu = 1$) to
$\approx 46$ for the large damping case ($k/\mu = 6$). The range of considered
values of the damping parameter $k/\mu$ is taken following estimates made in
Ref.~\cite{SDP} (see Fig.~10 of this Reference).
\begin{figure}[htb]
\includegraphics[width=1\columnwidth]{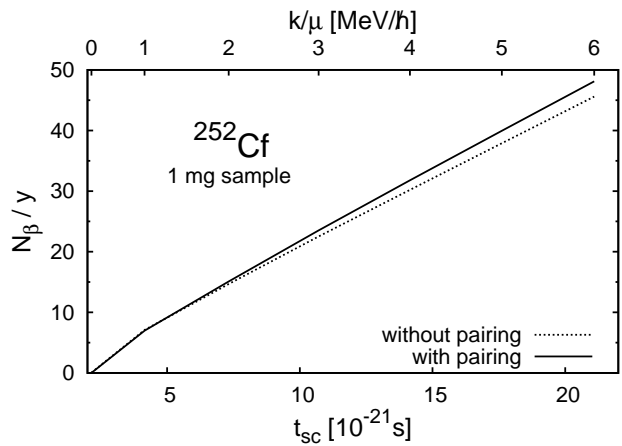}
\caption{Number of beta decays per year for a 1 mg sample of $^{252}$Cf as a
function of the descent time $t_{sc}$ from the exit to the scission points.}
\label{fig6}
\end{figure}

The previous figures are given for an evaluation which does not take into
account pairing correlations effects. As seen of Fig.~\ref{fig6}, including
these correlations increases slightly the above given numbers (only by $\approx
2 \%$ for $k/\mu = 6$), this correction being an increasing function of the
damping parameter $k/\mu$.

\section{Conclusions}

This paper aimed at pointing out the strong dependence of the beta-decay
stability as a function of the nuclear deformation. From state of the art
macroscopic-microscopic calculations it has been claimed that a very heavy
nucleus like $^{252}$Cf, while beta-stable as well-known in its ground state
becomes instable near the fission exit point. Through a very simple description
of the collective dynamics, the transition rates of the Fermi beta decay during
the spontaneous fission process of this nucleus up to the scission point, have
been calculated. They depend of course on the amount of damping of the fission
collective mode which regulates the excitation energy available at each time,
for such a decay. In so far as these rates could prove to be experimentally
reachable, they would provide a much needed source of informations on the
pre-scission dynamics, as e.g. the average pre-scission kinetic energy or the
value of the reduced friction parameter.
Taking or not into account pairing correlations is not an important factor.

As a result a relatively weak number of beta decay are expected from our results
during the descent from the exit to the scission points for standard values of
the damping parameter. The capacity of existing experimental devices to assess
this conclusion for tractable amounts of $^{252}$Cf is questionable. Improving
these so far very crude estimates by a better treatment of the static and
dynamic parts of these calculations as well as exploring this phenomenon for
other nuclei (involving possibly the inclusion of Gamow-Teller decays as well)
will be undertaken.

\section*{Acknowledgements}

This work was supported in part by the Polish National Science Centre under
Grant No.~2013/11/B/ST2/04087.
One of the author (P.Q.) gratefully acknowledges the support of the
Polish-French cooperation agreement COPIN 08-131 and the warm welcome extended
to him during a visit to the Department of Theoretical Physics at the Maria
Curie Sk{\l}odowska University.


\begin{thebibliography}{00}
\bibitem{LSD} K. Pomorski and J. Dudek, Phys. Rev. C {\bf 67}, 044316  (2003).
\bibitem{YFP} K.T.R. Davies and J.R. Nix, Phys. Rev. C {\bf 14}, 1977 (1976).
\bibitem{MASS} J. Randrup, S.E.Larsson, P. Moller, S.G. Nilsson, K. Pomorski,
        and A.Sobiczewski, Phys. Rev. C {\bf 13}, 229 (1976).
\bibitem{HAO} T.V. Hao, P. Quentin, and L. Bonneau, Phys. Rev. C {\bf 86},
        064307 (2012).
\bibitem{WUM} C.S. Wu and S.A. Moszkowski, {/it Beta Decay},
         Interscience monographs and texts in physics and astronomy, v. 16,
	Interscience Publishers, 1966
\bibitem{SDP} E. Strumberger, K. Dietrich, K. Pomorski, Nucl. Phys.
        {\bf A529}, 522 (1991).
\end{thebibliography}
\end{document}